\documentclass[preprint,11pt]{elsarticle}

\usepackage{amsmath,amssymb}
\usepackage{graphicx}
\usepackage{booktabs}
\usepackage{multirow}
\usepackage{array}
\usepackage{subcaption}
\usepackage{float}
\usepackage{xcolor}
\usepackage{url}
\usepackage{hyperref}

\journal{Knowledge-Based Systems}

\title{
Lightweight Resolution-Aware Audio Deepfake Detection via Cross-Scale Attention and Consistency Learning
}

\author[buet]{K.~A.~Shahriar\corref{cor1}\fnref{orcid1}}

\cortext[cor1]{Corresponding author. Email: kh.ashikshahriar@gmail.com}

\fntext[orcid1]{ORCID: \href{https://orcid.org/0009-0008-2571-7485}{0009-0008-2571-7485}}

\address[buet]{
Department of Electrical and Electronic Engineering (EEE),\\
Bangladesh University of Engineering and Technology (BUET),\\
Dhaka, Bangladesh
}

\begin{document}

\begin{frontmatter}

% =========================
% Abstract
% =========================
\begin{abstract}
Audio deepfake detection has become increasingly challenging due to rapid advances in speech synthesis and voice conversion technologies, particularly under channel distortions, replay attacks, and real-world recording conditions. This paper proposes a \emph{resolution-aware} audio deepfake detection framework that explicitly models and aligns multi-resolution spectral representations through cross-scale attention and consistency learning. Unlike conventional single-resolution or implicit feature-fusion approaches, the proposed method enforces agreement across complementary time--frequency scales. The proposed framework is evaluated on three representative benchmarks: ASVspoof 2019 (LA and PA), the Fake-or-Real (FoR) dataset, and the In-the-Wild Audio Deepfake dataset under a speaker-disjoint protocol. The method achieves near-perfect performance on ASVspoof LA (EER 0.16\%), strong robustness on ASVspoof PA (EER 5.09\%), FoR rerecorded audio (EER 4.54\%), and in-the-wild deepfakes (AUC 0.98, EER 4.81\%), significantly outperforming single-resolution and non-attention baselines under challenging conditions. The proposed model remains lightweight and efficient, requiring only 159k parameters and less than 1~GFLOP per inference, making it suitable for practical deployment. Comprehensive ablation studies confirm the critical contributions of cross-scale attention and consistency learning, while gradient-based interpretability analysis reveals that the model learns resolution-consistent and semantically meaningful spectral cues across diverse spoofing conditions. These results demonstrate that explicit cross-resolution modeling provides a principled, robust, and scalable foundation for next-generation audio deepfake detection systems.
\end{abstract}

% =========================
% Keywords
% =========================
\begin{highlights}
\item A lightweight resolution-aware framework is introduced for audio deepfake detection
\item Cross-scale attention enables adaptive fusion of multi-resolution spectral features
\item Consistency learning enforces resolution-invariant representations for robustness
\item Strong generalization is demonstrated on ASVspoof, FoR, and in-the-wild benchmarks
\item Interpretability analysis reveals meaningful and resolution-consistent detection cues
\end{highlights}


\end{frontmatter}

% =========================

\section{Introduction}

An \emph{audio deepfake} refers to synthetic or manipulated speech generated by machine learning models that imitate the acoustic characteristics, linguistic patterns, and vocal identity of a real speaker \cite{shaaban2023audio,dixit2023review,wang2024deepfake,chadha2021deepfake}. Unlike traditional text-to-speech (TTS) systems, modern audio deepfakes are produced using deep neural architectures such as neural vocoders \cite{sun2023ai,patel2023deepfake}, autoregressive generators \cite{morrison2021chunked,masood2023deepfakes}, and diffusion-based models \cite{bhagtani2024attribution}, enabling highly natural and speaker-specific speech generation \cite{liu2025review}. By conditioning these models on limited speech samples, an attacker can produce arbitrary utterances that closely resemble a target individual’s voice, often rendering human perception insufficient for reliable discrimination \cite{rabhi2024audio,muller2022attacker,bilika2023hello}.

The rapid evolution of speech synthesis and voice conversion technologies has significantly lowered the barrier to generating convincing audio deepfakes \cite{mubarak2023survey}. While these advances support beneficial applications in accessibility, speech restoration, and human–computer interaction \cite{danry2022ai}, they simultaneously introduce severe security and societal risks \cite{lee2024tug,amezaga2021availability}. Audio deepfakes can be exploited to bypass speaker verification systems, impersonate public figures, fabricate evidence, or conduct social engineering and financial fraud \cite{sareen2022threats,hery2024audio,mcuba2023effect}. The growing availability of open-source synthesis toolkits and commercial voice-cloning services further amplifies the urgency of developing reliable and generalizable detection mechanisms. Early research on audio deepfake and spoofing detection primarily focused on identifying artifacts introduced by synthesis pipelines \cite{li2022comparative,chakravarty2024lightweight,grinberg2025does,jellali2025pushing}. These approaches relied on handcrafted acoustic features, such as phase inconsistencies, spectral statistics, or cepstral coefficients, combined with classical classifiers \cite{bisogni2024acoustic,sandotra2024comprehensive,gohari2025audio}. While effective against early-generation synthetic speech, such methods were inherently limited by their reliance on manually designed features and assumptions about specific synthesis artifacts. As synthesis models improved, many of these handcrafted cues became increasingly subtle or absent \cite{dixit2023review,khanjani2021deep}.

The emergence of deep learning marked a paradigm shift in audio deepfake detection. Convolutional and attention-based neural networks enabled models to learn discriminative representations directly from time-frequency inputs such as spectrograms and Mel representations \cite{verma2025deepfake,kumar2022deepfakes,heidari2024deepfake,passos2024review,sharma2025systematic}. Large-scale benchmarks and community-driven challenges \cite{edwards2024review} further accelerated progress by standardizing evaluation protocols and datasets. On controlled benchmarks, modern detectors now achieve near-ceiling performance, often surpassing human-level accuracy. However, this apparent success masks a critical vulnerability. Detection systems trained on clean or curated datasets frequently degrade when exposed to replay attacks, channel distortions, compression artifacts, or real-world recording conditions \cite{muller2024harder,li2025measuring}. In practice, audio deepfakes are rarely encountered in pristine form; they are transmitted through microphones, speakers, telecommunication channels, and online platforms. The resulting distribution shift reveals that many existing detectors implicitly rely on dataset-specific artifacts or narrow spectral cues that do not generalize beyond the training conditions. A fundamental limitation underlying this issue lies in how time-frequency representations are modeled. Most existing approaches operate on a single spectral resolution or apply naive fusion strategies when multiple resolutions are used. Yet, different resolutions capture fundamentally different information: fine-grained representations emphasize short-term spectral irregularities, while coarser resolutions encode longer-term temporal dynamics and prosodic structure \cite{jones1989resolution,umapathy2010audio}. Treating these representations independently—or merging them without explicit alignment—can lead to inconsistent or brittle decision boundaries, particularly under channel and replay distortions.

In this work, it is argued that robust audio deepfake detection requires \emph{explicit cross-resolution modeling}. To this end, a resolution-aware detection framework was proposed that jointly learns from multiple spectral resolutions while enforcing coherent representation alignment. The proposed architecture introduces a cross-scale attention mechanism that dynamically fuses complementary resolution-specific features, enabling the model to adaptively emphasize informative cues depending on the acoustic conditions. Additionally, a consistency learning objective is employed to regularize predictions across resolutions, encouraging the model to focus on resolution-invariant characteristics rather than spurious artifacts. The proposed framework is evaluated on a diverse set of benchmark datasets spanning controlled synthesis attacks, replay-based distortions, and in-the-wild audio deepfakes, including speaker-disjoint evaluation protocols. Experimental results demonstrate that the method achieves near-ceiling performance on standard benchmarks while significantly improving robustness under challenging real-world conditions. Extensive ablation studies validate the contribution of each architectural component, and gradient-based interpretability analysis provides insights into the spectral cues leveraged across resolutions.

In summary, this work makes three key contributions. First, it introduces a resolution-aware audio deepfake detection framework that explicitly models and aligns multi-resolution spectral representations. Second, it demonstrates that cross-scale attention and consistency learning substantially enhance robustness under replayed and unconstrained conditions. Third, it provides both empirical and interpretability-based evidence that explicit cross-resolution modeling is a critical factor for building generalizable and trustworthy audio deepfake detection systems.

\section{Proposed Method}

\subsection{Theoretical Motivation}

Audio deepfake detection can be formulated as a binary classification problem, where an input speech signal is mapped to a label indicating whether it is bona fide or spoofed. Let $x(t)$ denote a raw audio waveform and $y \in \{0,1\}$ its corresponding label, where $0$ and $1$ represent bona fide and spoofed speech, respectively. Most deep learning-based detection systems operate by first transforming $x(t)$ into a time--frequency representation and subsequently learning a discriminative mapping through a neural network.

\begin{figure}[t]
    \centering
    \includegraphics[width=\linewidth]{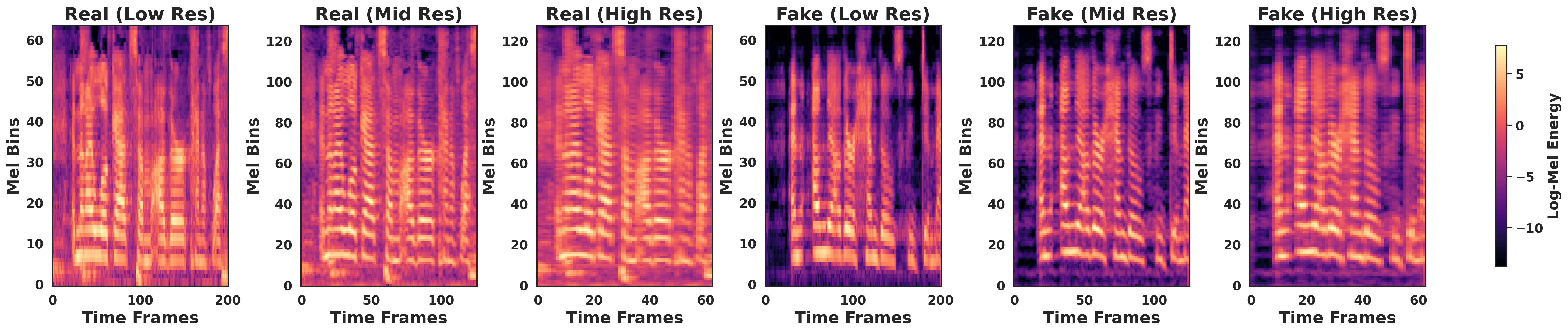}
    \caption{Multi-resolution log-Mel spectrograms of real and fake audio samples from the In-the-Wild dataset. Low-, mid-, and high-resolution representations reveal complementary temporal and spectral artifacts.}
    \label{fig:multi_res_tfr}
\end{figure}

A key observation motivating this work is that no single time--frequency resolution is sufficient to capture all artifacts introduced by modern speech synthesis and voice conversion systems. Fine-resolution spectrograms emphasize short-term spectral irregularities and high-frequency artifacts, whereas coarser resolutions encode longer-term temporal structures and prosodic inconsistencies. This phenomenon is illustrated in Fig.~\ref{fig:multi_res_tfr}, which shows multi-resolution log-Mel spectrograms of representative real and fake audio samples from the In-the-Wild dataset. While some synthesis artifacts are more pronounced at higher resolutions, others are only visible at coarser temporal scales, highlighting the complementary nature of different resolutions.

Formally, given an audio signal $x(t)$, a set of $K$ spectral representations can be extracted at different resolutions, denoted as
\begin{equation}
\mathcal{S}(x) = \{ S_1(x), S_2(x), \dots, S_K(x) \},
\end{equation}
where each $S_k(\cdot)$ corresponds to a time--frequency transformation with distinct temporal and spectral granularity. The proposed framework explicitly models and aligns these representations to exploit resolution-complementary cues for robust audio deepfake detection. Let $f_{\theta}(\cdot)$ denote a shared encoder network parameterized by $\theta$, which maps a spectral representation to a latent embedding space,
\begin{equation}
\mathbf{z}_k = f_{\theta}(S_k(x)) \in \mathbb{R}^{d}, \quad k = 1, \dots, K.
\end{equation}
In conventional multi-feature approaches, these embeddings are often concatenated or averaged before classification. However, such implicit fusion strategies do not explicitly model the relationships among resolutions and may allow resolution-specific artifacts to dominate the decision function. This can result in brittle representations that fail to generalize under channel distortions or rerecording conditions \cite{firc2025evaluation}.

To address this limitation, the proposed framework introduces an explicit cross-scale interaction mechanism that allows embeddings from different resolutions to attend to one another. Let $\mathbf{Z} = [\mathbf{z}_1, \mathbf{z}_2, \dots, \mathbf{z}_K]^\top \in \mathbb{R}^{K \times d}$ denote the stacked multi-resolution embeddings. Cross-scale attention is applied as
\begin{equation}
\tilde{\mathbf{Z}} = \mathrm{Attn}(\mathbf{Z}, \mathbf{Z}, \mathbf{Z}),
\end{equation}
where $\mathrm{Attn}(\cdot)$ denotes a multi-head self-attention operator. This formulation enables each resolution-specific embedding to dynamically reweight information from other resolutions, thereby capturing complementary spectral cues. While cross-scale attention facilitates feature fusion, it does not explicitly enforce consistency across resolutions. In practice, embeddings extracted from different resolutions may encode conflicting or resolution-dependent cues, particularly under channel distortions. To mitigate this effect, a consistency learning objective is introduced to encourage alignment among embeddings corresponding to bona fide speech. Specifically, the normalized embeddings $\hat{\mathbf{z}}_k = \mathbf{z}_k / \lVert \mathbf{z}_k \rVert_2$ are encouraged to be similar across resolutions,
\begin{equation}
\mathcal{L}_{\text{cons}} = \sum_{i < j} \mathbb{E}_{x \sim \mathcal{D}_{\text{real}}}
\left[ \lVert \hat{\mathbf{z}}_i - \hat{\mathbf{z}}_j \rVert_2^2 \right],
\end{equation}
where $\mathcal{D}_{\text{real}}$ denotes the distribution of bona fide speech. This objective reflects the assumption that genuine speech should exhibit resolution-invariant characteristics, whereas spoofed speech often introduces inconsistencies across spectral scales.

The final classification score is obtained by aggregating the attended embeddings and applying a linear classifier,
\begin{equation}
\hat{y} = \sigma\left( g\left( \frac{1}{K} \sum_{k=1}^{K} \tilde{\mathbf{z}}_k \right) \right),
\end{equation}
where $g(\cdot)$ denotes a learnable affine transformation and $\sigma(\cdot)$ is the sigmoid function. The overall training objective combines the binary cross-entropy loss $\mathcal{L}_{\text{cls}}$ with the consistency regularization term,
\begin{equation}
\mathcal{L} = \mathcal{L}_{\text{cls}} + \lambda \mathcal{L}_{\text{cons}},
\end{equation}
where $\lambda$ controls the contribution of consistency learning. This formulation explicitly encodes cross-resolution relationships and enforces resolution-consistent representations, providing a principled mechanism to improve robustness and generalization in audio deepfake detection.

\subsection{Datasets}

The proposed method is evaluated on three publicly available audio deepfake detection datasets that collectively cover controlled, rerecorded, and real-world conditions. These datasets are widely used in the literature and provide complementary evaluation scenarios for assessing robustness and generalization.

\subsubsection{ASVspoof 2019 Dataset}

The ASVspoof 2019 dataset \cite{wang2020asvspoof} is a widely used benchmark for audio spoofing and deepfake detection and consists of two complementary evaluation scenarios: Logical Access (LA) and Physical Access (PA). Both subsets are provided at a sampling rate of 16~kHz and follow standardized experimental protocols. The LA subset contains bona fide speech recordings and spoofed utterances generated using a diverse set of TTS and VC systems. This subset represents a controlled experimental setting in which spoofing artifacts are primarily algorithmic rather than environmental. The PA subset, in contrast, focuses on replay-based spoofing attacks, where bona fide and spoofed speech are captured through physical recording and playback devices under varying acoustic environments and channel conditions, thereby introducing realistic environmental distortions.

The dataset is partitioned into training, development, and evaluation subsets following a predefined protocol. In this work, experiments are conducted on both the LA and PA subsets to assess the proposed method under algorithmic and physical spoofing scenarios. The training and development subsets are used for model optimization and validation, respectively, following established evaluation practices in the literature. A summary of the dataset statistics used in this study is provided in Table~\ref{tab:asvspoof_stats}.

\begin{table}[t]
\centering
\caption{Summary of ASVspoof 2019 LA and PA dataset statistics used in this study }
\label{tab:asvspoof_stats}
\begin{tabular}{lcccc}
\toprule
\textbf{Subset} & \textbf{Split} & \textbf{Spoof} & \textbf{Bonafide} & \textbf{Total} \\
\midrule
\multirow{2}{*}{LA} 
 & Train & 22{,}800 & 2{,}580 & 25{,}380 \\
 & Dev   & 22{,}296 & 2{,}548 & 24{,}844 \\
\midrule
\multirow{2}{*}{PA} 
 & Train & 48{,}600 & 5{,}400 & 54{,}000 \\
 & Dev   & 24{,}300 & 5{,}400 & 29{,}700 \\
\bottomrule
\end{tabular}
\end{table}

\subsubsection{Fake-or-Real (FoR) Dataset}

The Fake-or-Real (FoR) dataset \cite{8906599} is a large-scale benchmark designed for synthetic speech and audio deepfake detection. It contains over 198{,}000 utterances comprising both bona fide human speech and spoofed speech generated by a wide range of modern TTS systems. Unlike earlier datasets based on classical synthesis techniques, FoR focuses on DL-based speech synthesis methods, including both open-source and commercial systems. The spoofed portion of the dataset includes utterances generated using state-of-the-art neural TTS models such as DeepVoice~3, Google Cloud TTS, WaveNet, Amazon AWS Polly, Microsoft Azure TTS, and Baidu Cloud TTS, covering a total of 33 synthetic voices. Bona fide speech samples are collected from diverse sources, including open speech corpora and real-world recordings, ensuring variability in speakers, accents, microphones, and recording environments.

The FoR dataset is released in multiple versions. The \textit{for-original} version contains the raw collected utterances, while the \textit{for-norm} version applies normalization steps including resampling to 16~kHz, volume normalization, mono conversion, and silence trimming. The \textit{for-2sec} version further truncates utterances to a fixed duration of two seconds to eliminate length-related bias. Finally, the \textit{for-rerecorded} version simulates replay attacks by replaying and re-recording utterances using consumer-grade playback and recording devices, introducing realistic channel distortions. In this work, the normalized, two-second, and rerecorded versions of the FoR dataset are used to evaluate the proposed resolution-aware framework under controlled, normalized, and replay-like conditions. These subsets enable a systematic assessment of robustness to recording variability and channel effects.

\begin{table}[t]
\centering
\caption{Data distribution of the FoR dataset subsets used in this study (16~kHz sampling rate).}
\label{tab:for_stats}
\begin{tabular}{lccc}
\toprule
\textbf{Subset} & \textbf{Spoof} & \textbf{Bonafide} & \textbf{Total} \\
\midrule
for-norm       & 34{,}700 & 34{,}700 & 69{,}400 \\
for-2sec       & 8{,}935  & 8{,}935  & 17{,}870 \\
for-rerecorded & 8{,}935  & 8{,}935  & 17{,}870 \\
\bottomrule
\end{tabular}
\end{table}

\subsubsection{In-the-Wild Audio Deepfake Dataset}

To evaluate real-world generalization, experiments are conducted on the In-the-Wild Audio Deepfake dataset \cite{muller2022does}. It contains bona fide and spoofed speech collected from publicly available online sources, including social media platforms and video streaming websites, resulting in substantial variability in speakers, recording environments, compression artifacts, and background noise. The dataset comprises audio recordings from 58 public figures, including politicians and other well-known individuals. Both bona fide and spoofed audio samples are provided for each speaker, enabling speaker-aware and speaker-disjoint evaluation protocols. In total, the dataset contains approximately 20.8 hours of bona fide speech and 17.2 hours of spoofed speech, corresponding to an average of 23 minutes of bona fide audio and 18 minutes of spoofed audio per speaker. All audio samples are provided at a sampling rate of 16 kHz.

In this study, a speaker-disjoint splitting protocol is adopted to ensure that speakers in the training, validation, and test sets do not overlap. This protocol prevents identity leakage and provides a rigorous evaluation of the model’s ability to generalize to unseen speakers and recording conditions. Owing to its diverse and unconstrained nature, this dataset presents a significantly more challenging evaluation scenario than curated benchmarks and serves as a realistic proxy for deployment conditions. A summary of the dataset statistics used in this work is reported in Table~\ref{tab:inthewild_stats}.

\begin{table}[t]
\centering
\caption{Summary of the In-the-Wild Audio Deepfake dataset used in this study.}
\label{tab:inthewild_stats}
\begin{tabular}{lc}
\toprule
\textbf{Characteristic} & \textbf{Value} \\
\midrule
Total samples        & 31{,}779 \\
Bona fide samples    & 19{,}963 \\
Spoofed samples      & 11{,}816 \\
Number of speakers   & 58 \\
Total bona fide audio duration & 20.8 hours \\
Total spoofed audio duration   & 17.2 hours \\
Sampling rate        & 16~kHz \\
\bottomrule
\end{tabular}
\end{table}

\subsection{Preprocessing and Feature Extraction}
A unified preprocessing and feature extraction pipeline is adopted for all experiments to ensure consistent and fair evaluation across datasets with diverse recording conditions and durations. This pipeline is designed to preserve relevant spectral artifacts while minimizing confounding factors such as sampling rate variability and utterance length differences.

\subsubsection{Audio Preprocessing}

All audio signals are resampled to a common sampling rate of 16~kHz and converted to single-channel (mono) format. For datasets containing variable-length utterances, fixed-duration segments are extracted to reduce bias associated with utterance length. For the FoR dataset, segments of 2~seconds are obtained using random cropping during preprocessing. For the In-the-Wild Audio Deepfake dataset, longer segments of 4~seconds are extracted using random cropping during training, while padding is applied when necessary for shorter recordings. Random cropping is applied only during training to introduce temporal variability and improve generalization \cite{li2003audio}, whereas deterministic cropping or full segments are used during validation.

\subsubsection{Multi-Resolution Spectral Feature Extraction}

Multiple log-mel spectrogram representations are extracted from each audio segment to capture complementary spectral representations. Specifically, three mel-spectrograms are computed using distinct short-time Fourier transform (STFT) configurations:
\begin{itemize}
\item A fine-resolution representation with $n_{\text{fft}}=400$, hop length of 160 samples, and 64 mel frequency bins.
\item A medium-resolution representation with $n_{\text{fft}}=1024$, hop length of 256 samples, and 128 mel frequency bins.
\item A coarse-resolution representation with $n_{\text{fft}}=2048$, hop length of 512 samples, and 128 mel frequency bins.
\end{itemize}

Log-amplitude scaling is applied to each mel-spectrogram to stabilize optimization and emphasize perceptually relevant spectral differences. The resulting representations provide complementary views of short-term spectral irregularities and longer-term temporal patterns, which are jointly exploited by the proposed resolution-aware framework.

\subsection{Model Architecture}

The proposed resolution-aware audio deepfake detection model is designed to jointly exploit complementary spectral resolutions while maintaining computational efficiency. The architecture consists of three main components: a shared convolutional encoder, a cross-scale attention module, and resolution-agnostic classification heads. An overview of the model complexity is summarized in Table~\ref{tab:model_complexity}.

\subsubsection{Shared Encoder}

Each input audio segment is represented by three log-mel spectrograms computed at different time--frequency resolutions. A shared convolutional encoder is applied independently to each resolution, ensuring parameter efficiency and consistent feature extraction across scales. The encoder consists of three convolutional layers with ReLU activations, followed by adaptive average pooling to produce fixed-dimensional embeddings. Given an input spectrogram $S_k$, the encoder maps it to a latent representation $\mathbf{z}_k \in \mathbb{R}^{128}$.

For experiments involving dataset-aware training (FoR), learnable scale-and-shift parameters ($\gamma$ and $\beta$) are applied to the latent embeddings to allow lightweight dataset-conditioned modulation. This operation preserves the shared encoder structure while enabling minor distribution-specific adaptation.

\subsubsection{Cross-Scale Attention}

Three latent vectors are stacked and passed to a multi-head self-attention module. This cross-scale attention mechanism enables each resolution to attend to complementary information from other resolutions, allowing the model to dynamically emphasize resolution-invariant and discriminative spectral cues. The attention output is aggregated by mean pooling across resolutions, producing a fused representation that integrates multi-scale information.

\subsubsection{Classification Head}

The fused embedding is passed to a lightweight linear classifier that outputs a scalar logit corresponding to the spoofing probability. For multi-dataset experiments, separate classification heads are used while sharing the encoder and attention modules. This design isolates dataset-specific decision boundaries without increasing the representational capacity of the shared feature extractor.

\subsubsection{Model Complexity}

The proposed architecture remains compact and computationally efficient. The total number of trainable parameters is approximately $1.6 \times 10^5$, with a model size below 1~MB (0.62 MB) in single-precision format.

\begin{table}[t]
\centering
\caption{Parameter count and computational complexity of the proposed model.}
\label{tab:model_complexity}
\begin{tabular}{l r}
\toprule
\textbf{Module} & \textbf{\# Parameters} \\
\midrule
Encoder Conv Layer 1 & 320 \\
Encoder Conv Layer 2 & 18{,}496 \\
Encoder Conv Layer 3 & 73{,}856 \\
Dataset Modulation ($\gamma$) & 384 \\
Dataset Modulation ($\beta$) & 384 \\
Multi-head Attention & 49{,}536 \\
Attention Output Projection & 16{,}512 \\
Classifier Head (per dataset) & 129 \\
\midrule
Total Trainable Parameters & 159{,}875 \\
\bottomrule
\end{tabular}
\end{table}

The proposed model contains 159{,}875 trainable parameters, corresponding to a memory footprint of approximately 0.62~MB in single-precision and 0.30~MB in half-precision format. The total number of floating-point operations per forward pass is approximately 936.66~MFLOPs, while the peak activation memory usage during training is 42.07~MB. These characteristics demonstrate that the model achieves strong detection performance while maintaining a favorable trade-off between accuracy and computational efficiency.

\subsection{Model Training and Evaluation}

This section describes the training procedure and evaluation methodology used to assess the proposed resolution-aware audio deepfake detection framework. All experiments are conducted using a unified training strategy to ensure fair comparison across datasets and experimental conditions.

\subsubsection{Training Procedure}

The proposed model is trained as a binary classifier to distinguish between bona fide and spoofed speech. Given an input audio segment represented by multiple spectral resolutions, the model outputs a scalar logit indicating the likelihood of spoofing. Training is performed using the binary cross-entropy (BCE) loss. For experiments involving consistency learning, an additional regularization term is incorporated to encourage alignment among resolution-specific embeddings, and the total loss is computed as a weighted sum of classification and consistency losses.

All models are optimized using the Adam optimizer with a fixed learning rate of $1\times10^{-4}$. Mini-batch training is employed with a batch size of 32. Training is conducted for 15 epochs, with the best-performing model selected based on validation performance.

\subsubsection{Model Selection}

Model selection is performed using the validation sets corresponding to each dataset. The equal error rate (EER) is used as the primary selection criterion, as it provides a threshold-independent measure that is widely adopted in the audio spoofing literature. For each experiment, the model achieving the lowest validation EER is selected and saved for subsequent evaluation.

\subsubsection{Evaluation Metrics}

Performance is evaluated using multiple complementary metrics. Classification accuracy is reported to provide an intuitive measure of correctness. The EER is used as the primary metric to quantify the trade-off between false acceptance and false rejection rates. In addition, the area under the receiver operating characteristic curve (ROC-AUC) is reported for selected experiments to characterize the overall discriminative capability of the model independent of the decision threshold.

For the ASVspoof 2019 LA and PA datasets, evaluation follows the official protocol, with training conducted on the training subset and performance assessed on the development subset. For the FoR dataset, evaluation is conducted separately on the normalized, two-second, and rerecorded versions to analyze robustness under increasing levels of channel distortion. For the In-the-Wild Audio Deepfake dataset, a speaker-disjoint protocol is employed, ensuring that speakers in the training, validation, and test sets do not overlap.

\begin{figure*}[t]
    \centering
    \includegraphics[width=\textwidth]{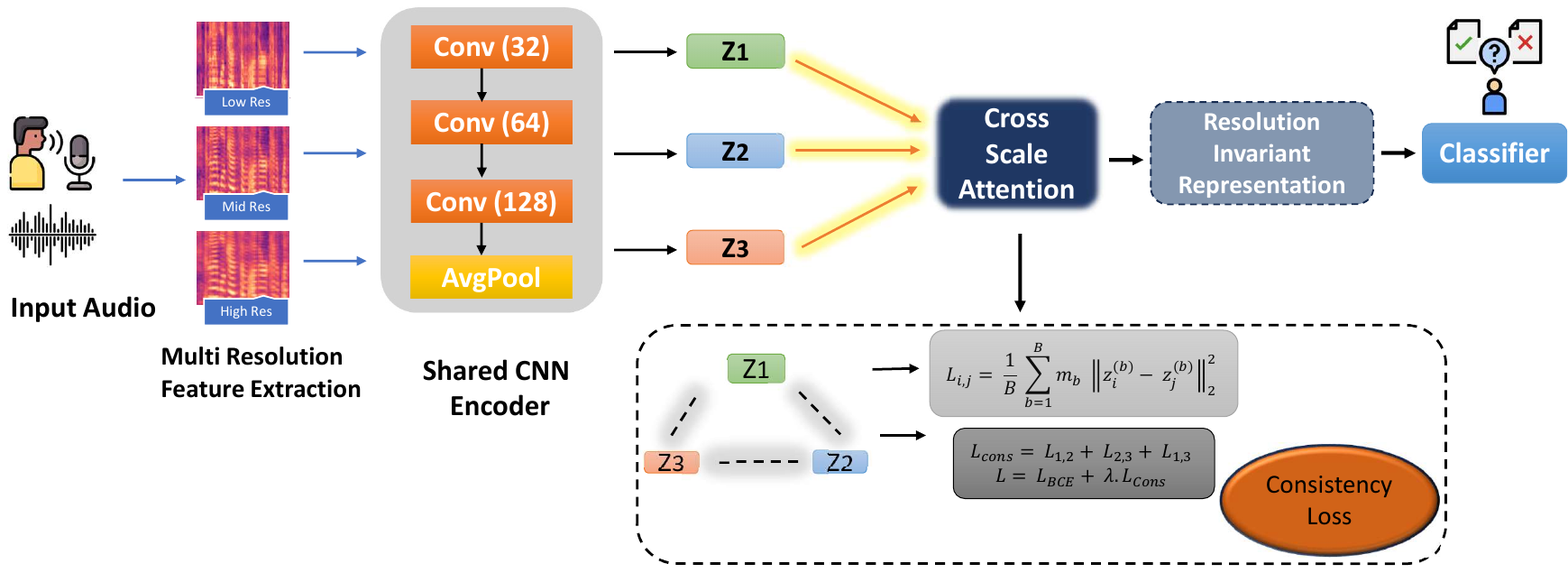}
    \caption{Overview of the proposed resolution-aware audio deepfake detection framework. An input audio signal is transformed into low-, mid-, and high-resolution Mel spectrograms, which are processed by a shared convolutional encoder to extract resolution-specific embeddings $z_1$, $z_2$, and $z_3$. A cross-scale attention module dynamically fuses these embeddings to form a resolution-invariant representation for classification. During training, a cross-resolution consistency loss is applied to bona fide speech to enforce alignment across resolutions, improving robustness under replay and real-world conditions.}
    \label{fig:methodology}
\end{figure*}

As illustrated in Fig.~\ref{fig:methodology}, the proposed framework jointly learns from multiple spectral resolutions through a shared encoder, cross-scale attention, and consistency regularization.

% -------------------------------------------------

\section{Results and Analysis}

\subsection{Results on ASVspoof 2019}

\subsubsection{Logical Access (LA) Results}

Table~\ref{tab:asvspoof_la_results} summarizes the classification performance on the ASVspoof 2019 LA development set. The proposed method achieves near-perfect performance, with an accuracy of 99.90\%, an AUC of 1.000, and an EER of 0.0016. Both bona fide and spoofed classes are detected with extremely high precision and recall, indicating that the model effectively captures algorithmic artifacts introduced by TTS and VC systems.

\begin{table}[t]
\centering
\caption{Performance on ASVspoof 2019 LA (Development Set).}
\label{tab:asvspoof_la_results}
\begin{tabular}{lccc}
\toprule
\textbf{Class} & \textbf{Precision} & \textbf{Recall} & \textbf{F1-score} \\
\midrule
Bona fide & 0.9933 & 0.9965 & 0.9949 \\
Spoof & 0.9996 & 0.9992 & 0.9994 \\
\midrule
Accuracy & \multicolumn{3}{c}{0.9990} \\
AUC & \multicolumn{3}{c}{1.0000} \\
EER & \multicolumn{3}{c}{0.0016} \\
\bottomrule
\end{tabular}
\end{table}

\begin{figure*}[t]
\centering
\includegraphics[width=\textwidth]{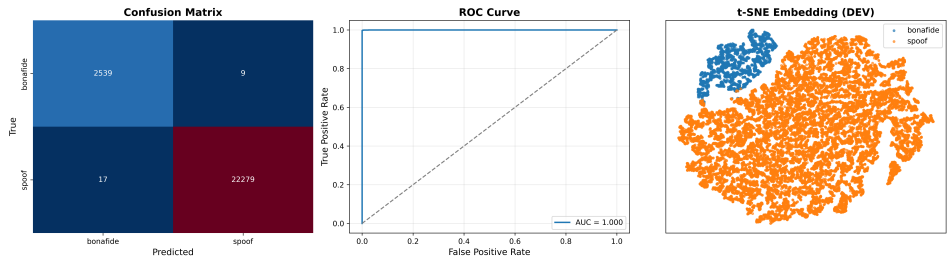}
\caption{ASVspoof 2019 LA results: confusion matrix (left), ROC curve (center), and t-SNE embedding of development set representations (right).}
\label{fig:asvspoof_la_vis}
\end{figure*}

Figure~\ref{fig:asvspoof_la_vis} illustrates the confusion matrix, ROC curve, and t-SNE embedding for the LA subset. The confusion matrix shows only a negligible number of misclassifications, while the ROC curve demonstrates near-ideal separability between classes. The t-SNE visualization further reveals a clear clustering structure, with bona fide and spoofed samples forming well-separated manifolds in the learned embedding space.

\subsubsection{Physical Access (PA) Results}

Table~\ref{tab:asvspoof_pa_results} reports performance on the ASVspoof 2019 PA development set. In this more challenging replay-based scenario, the proposed model achieves an accuracy of 95.32\%, an AUC of 0.9884, and an EER of 0.0509. While performance remains strong, the increased error rate compared to the LA subset reflects the greater variability introduced by acoustic environments, playback devices, and recording channels.

\begin{table}[t]
\centering
\caption{Performance on ASVspoof 2019 PA (Development Set).}
\label{tab:asvspoof_pa_results}
\begin{tabular}{lccc}
\toprule
\textbf{Class} & \textbf{Precision} & \textbf{Recall} & \textbf{F1-score} \\
\midrule
Bona fide & 0.8647 & 0.8804 & 0.8725 \\
Spoof & 0.9733 & 0.9694 & 0.9713 \\
\midrule
Accuracy & \multicolumn{3}{c}{0.9532} \\
AUC & \multicolumn{3}{c}{0.9884} \\
EER & \multicolumn{3}{c}{0.0509} \\
\bottomrule
\end{tabular}
\end{table}

\begin{figure*}[t]
\centering
\includegraphics[width=\textwidth]{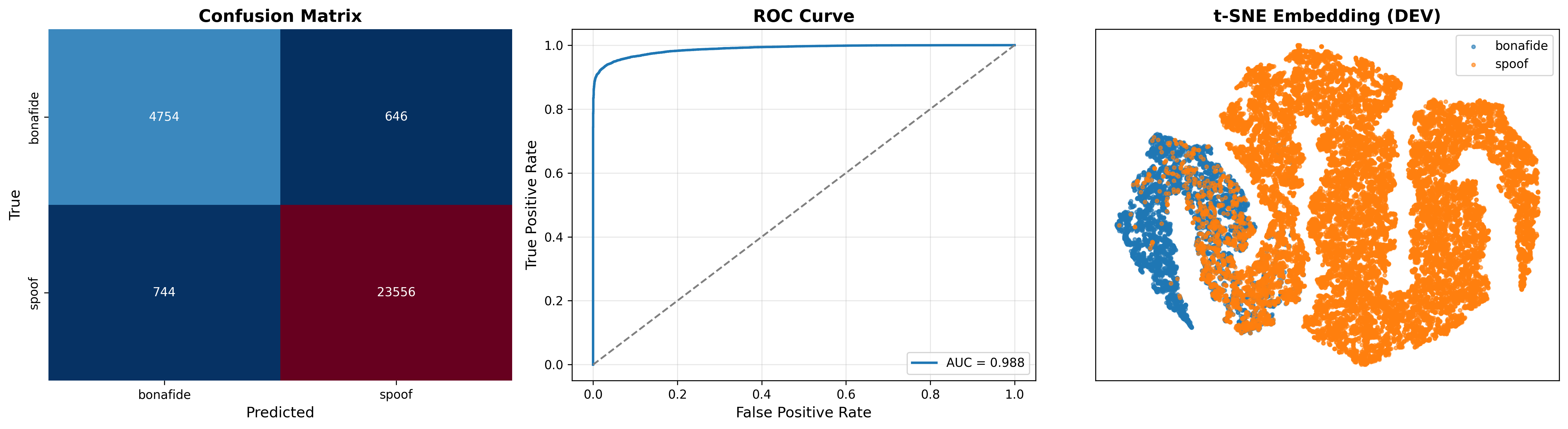}
\caption{ASVspoof 2019 PA results: confusion matrix (left), ROC curve (center), and t-SNE embedding of development set representations (right).}
\label{fig:asvspoof_pa_vis}
\end{figure*}

Figure~\ref{fig:asvspoof_pa_vis} presents the corresponding confusion matrix, ROC curve, and t-SNE embedding for the PA subset. Compared to the LA case, the confusion matrix indicates a higher rate of bona fide samples misclassified as spoofed, which is consistent with replay attacks introducing channel distortions that partially resemble spoofing artifacts. The ROC curve nevertheless maintains a steep ascent, indicating strong discriminative capability. The t-SNE visualization shows increased overlap between classes, reflecting the inherent difficulty of replay attack detection, yet still preserves a meaningful degree of separation in the learned representation space.

\subsection{Results on the FoR Dataset}

Table~\ref{tab:for_results} summarizes the classification performance across all FoR subsets. The proposed method achieves consistently strong results, with near-perfect performance on normalized and duration-controlled data and graceful degradation under rerecorded conditions. In particular, the \textit{for-2sec} subset yields the highest overall accuracy of 99.12\% with an EER of 0.0057.
\begin{table}[t]
\centering
\caption{Performance on the FoR dataset across different recording conditions.}
\label{tab:for_results}
\begin{tabular}{lccc}
\toprule
\textbf{Subset} & \textbf{Accuracy} & \textbf{EER} & \textbf{ROC-AUC} \\
\midrule
for-2sec & 0.9912 & 0.0057 & 0.9993 \\
for-norm & 0.9810 & 0.0046 & 0.9996 \\
for-rerec & 0.9563 & 0.0454 & 0.9900 \\
\bottomrule
\end{tabular}
\end{table}

\subsubsection{Results on for-2sec}

\begin{figure*}[t]
\centering
\includegraphics[width=\textwidth]{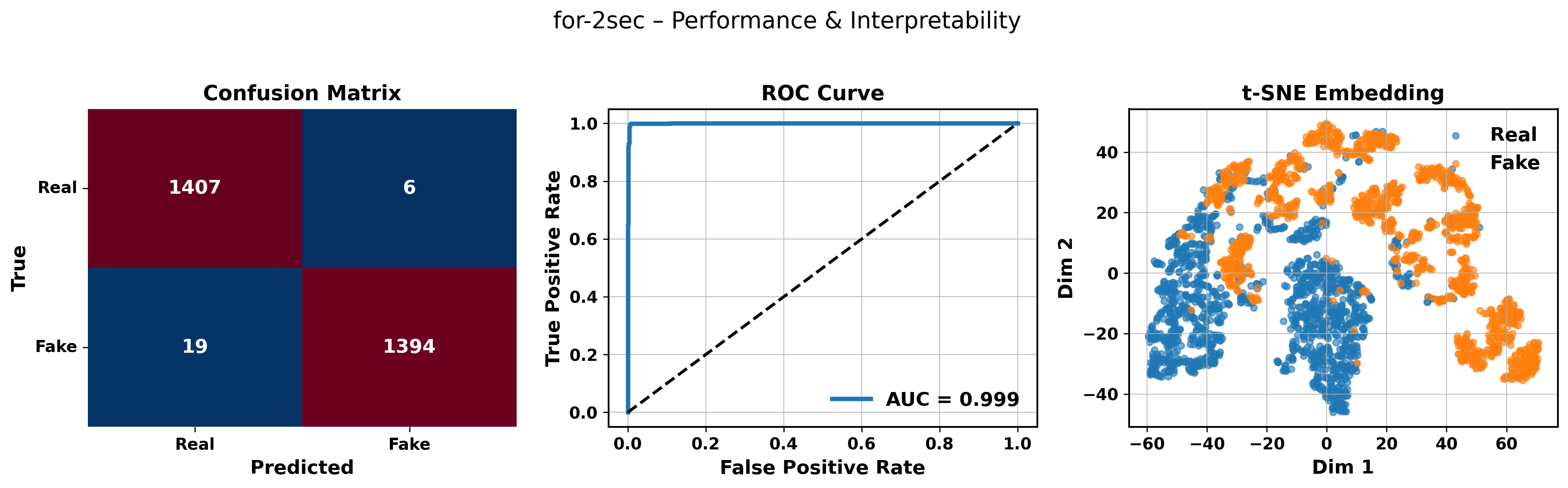}
\caption{FoR for-2sec results: confusion matrix (left), ROC curve (center), and t-SNE embedding (right).}
\label{fig:for_2sec_vis}
\end{figure*}

Figure~\ref{fig:for_2sec_vis} illustrates the confusion matrix, ROC curve, and t-SNE embedding for the \textit{for-2sec} subset. The confusion matrix shows only a small number of misclassifications, with balanced error rates across classes. The ROC curve exhibits near-ideal behavior, achieving an AUC of 0.9993. The t-SNE visualization reveals two well-separated clusters corresponding to bona fide and spoofed speech, indicating that the learned embeddings remain highly discriminative even when utterances are restricted to a fixed duration of two seconds.

\subsubsection{Results on for-norm}

\begin{figure*}[t]
\centering
\includegraphics[width=\textwidth]{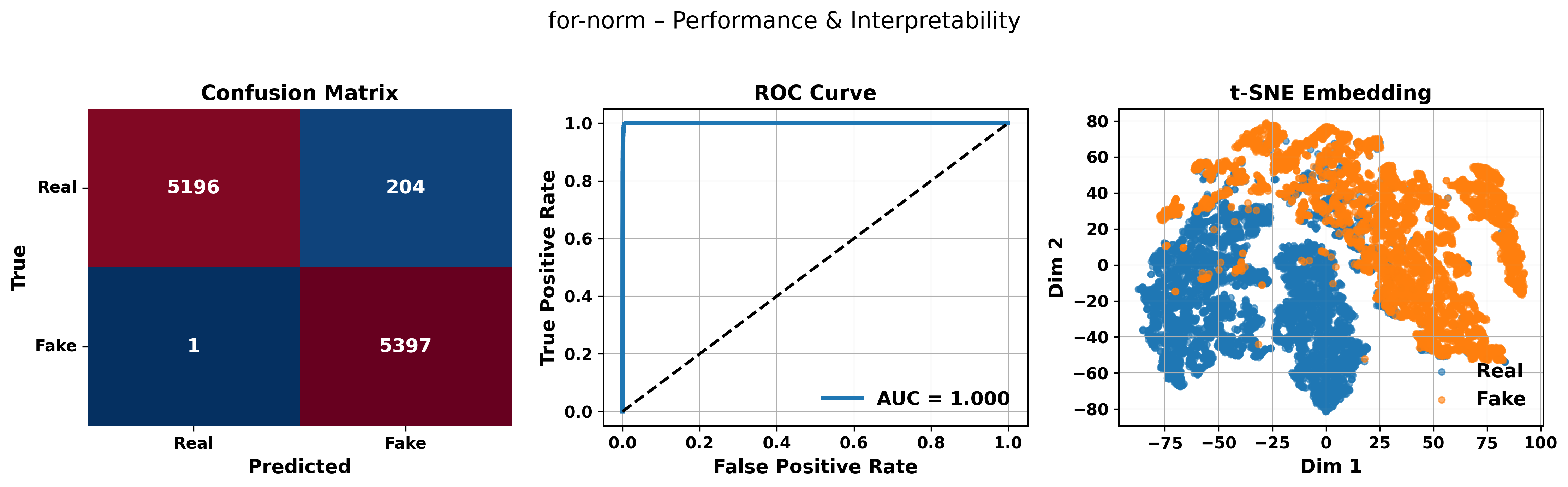}
\caption{FoR for-norm results: confusion matrix (left), ROC curve (center), and t-SNE embedding (right).}
\label{fig:for_norm_vis}
\end{figure*}

The results on the normalized FoR subset are presented in Figure~\ref{fig:for_norm_vis}. This setting represents a clean and controlled scenario in which recording conditions are standardized. The proposed method achieves an accuracy of 98.10\% and an AUC of 0.9996. The confusion matrix indicates extremely low false-positive and false-negative rates, while the t-SNE embedding shows clear separation between real and fake speech manifolds.

\subsubsection{Results on for-rerec}

\begin{figure*}[t]
\centering
\includegraphics[width=\textwidth]{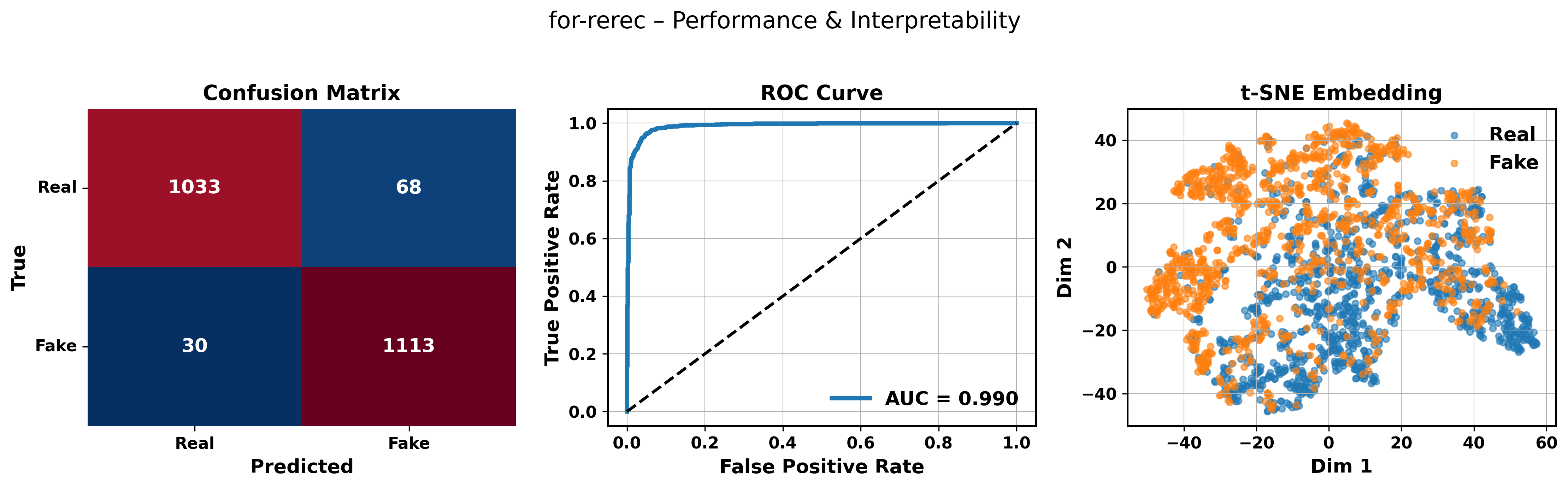}
\caption{FoR for-rerec results: confusion matrix (left), ROC curve (center), and t-SNE embedding (right).}
\label{fig:for_rerec_vis}
\end{figure*}

Figure~\ref{fig:for_rerec_vis} reports results on the rerecorded FoR subset, which simulates replay attacks and channel distortions. While performance decreases compared to the clean subsets, the proposed method still achieves a strong accuracy of 95.63\% with an AUC of 0.9900. The confusion matrix reveals a moderate increase in misclassification of bona fide samples, reflecting the impact of replay-induced artifacts. The t-SNE visualization shows increased overlap between classes, yet preserves a discernible separation structure, indicating that the learned representations retain meaningful spoofing-related information even under adverse recording conditions.

\subsection{Results on In-the-Wild Audio Deepfakes}

Table~\ref{tab:inthewild_results} summarizes the test-set performance of the proposed model. The method achieves an overall accuracy of 95.70\%, an ROC-AUC of 0.9800, and an EER of 0.0481.
\begin{table}[t]
\centering
\caption{Performance on the In-the-Wild Audio Deepfake Dataset (Test Set).}
\label{tab:inthewild_results}
\begin{tabular}{lccc}
\toprule
\textbf{Metric} & \textbf{Value} \\
\midrule
Accuracy & 0.9570 \\
ROC-AUC & 0.9800 \\
EER & 0.0481 \\
\bottomrule
\end{tabular}
\end{table}

\begin{figure}[t]
\centering
\includegraphics[width=0.95\linewidth]{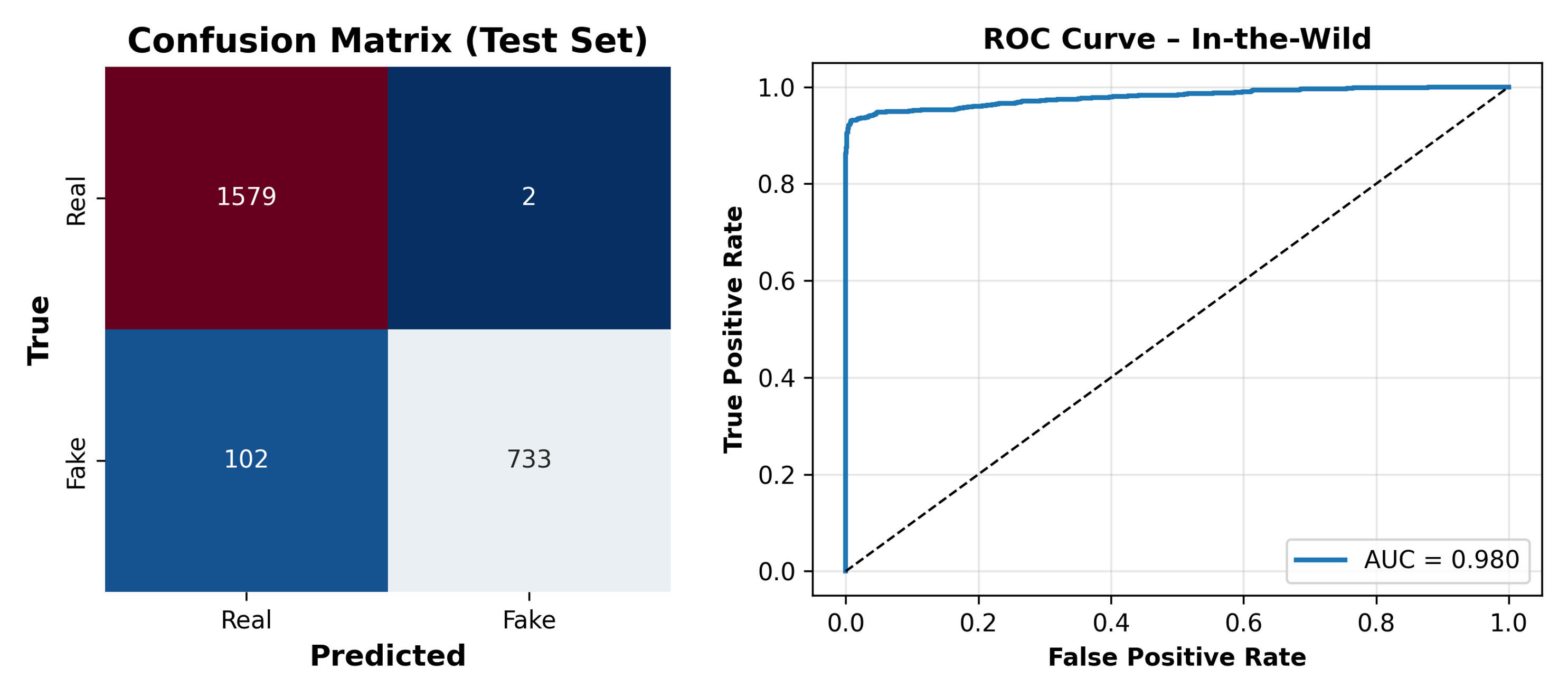}
\caption{In-the-Wild Audio Deepfake results on the test set: confusion matrix (left) and ROC curve (right).}
\label{fig:inthewild_vis}
\end{figure}

Figure~\ref{fig:inthewild_vis} presents the confusion matrix and ROC curve for the test set. The confusion matrix reveals that the vast majority of bona fide utterances are correctly classified, with only two false-positive errors, resulting in a near-perfect recall of 1.00 for the real class. Spoofed samples are detected with high precision, although a moderate number of false negatives are observed. The ROC curve exhibits a steep ascent near the origin, indicating strong discriminative capability across a wide range of thresholds. The achieved AUC of 0.98 confirms that the learned representations remain robust under real-world acoustic conditions and speaker variability.

\subsection{Interpretability Analysis}

An interpretability analysis is conducted using multi-resolution time--frequency visualizations and Grad-CAM–based activation mapping. The analysis focuses on three representative evaluation scenarios: (i) ASVspoof 2019 (PA), (ii) the re-recorded subset of the FoR dataset, and (iii) the In-the-Wild audio deepfake dataset.

\subsubsection{Grad-CAM Analysis on ASVspoof PA}

\begin{figure}[t]
    \centering
    \includegraphics[width=\linewidth]{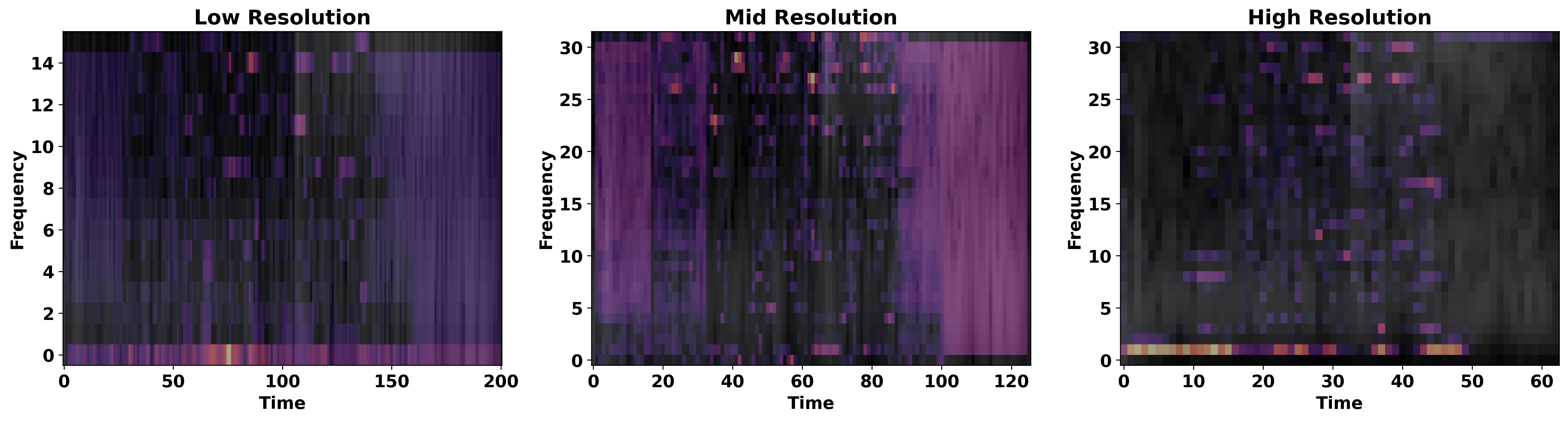}
    \caption{Resolution-wise saliency visualization on the ASVspoof 2019 PA dataset. From left to right: low-, mid-, and high-resolution Mel-spectrogram representations.}
    \label{fig:interp_asvspoof}
\end{figure}

Figure~\ref{fig:interp_asvspoof} illustrates the saliency distributions across low-, mid-, and high-resolution Mel-spectrogram representations for the ASVspoof 2019 PA dataset. The low-resolution representation primarily captures long-term temporal energy distributions and coarse spectral envelopes, which provide contextual cues related to speech rhythm and prosody. Mid-resolution features emphasize formant transitions and temporal modulation patterns, while high-resolution representations highlight fine-grained spectral irregularities, particularly in higher frequency regions.

\subsubsection{Grad-CAM Analysis on Re-recorded Audio (FoR)}

\begin{figure}[t]
    \centering
    \includegraphics[width=\linewidth]{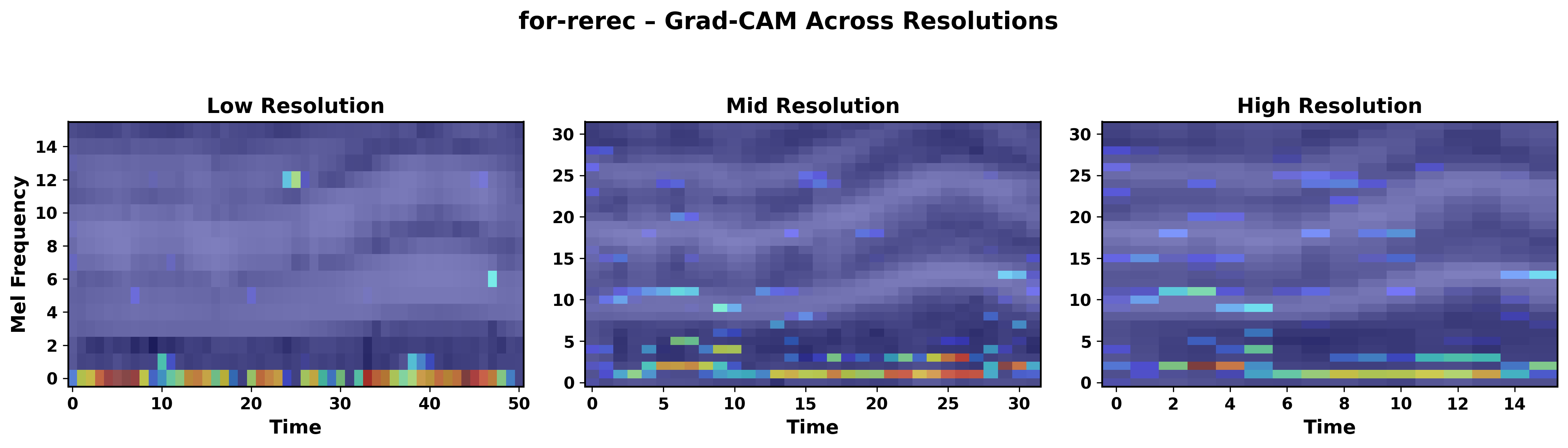}
    \caption{Grad-CAM visualization across resolutions for the FoR re-recorded dataset, illustrating attention shifts induced by replay and channel effects.}
    \label{fig:interp_for}
\end{figure}

Figure~\ref{fig:interp_for} presents Grad-CAM activation maps across resolutions for the FoR re-recorded subset, which simulates replay attacks and channel distortions. Compared to the controlled PA condition, the activation patterns are more spatially dispersed, with increased emphasis on mid- and high-frequency regions during transient segments. This behavior suggests that the model adapts its attention to artifacts introduced by the playback--recording chain, including microphone coloration, reverberation, and spectral smearing. Importantly, the low-resolution pathway continues to contribute stable contextual information, while higher resolutions capture channel-induced inconsistencies.

\subsubsection{Interpretability in Real-World Conditions (In-the-Wild)}

\begin{figure}[t]
    \centering
    \includegraphics[width=\linewidth]{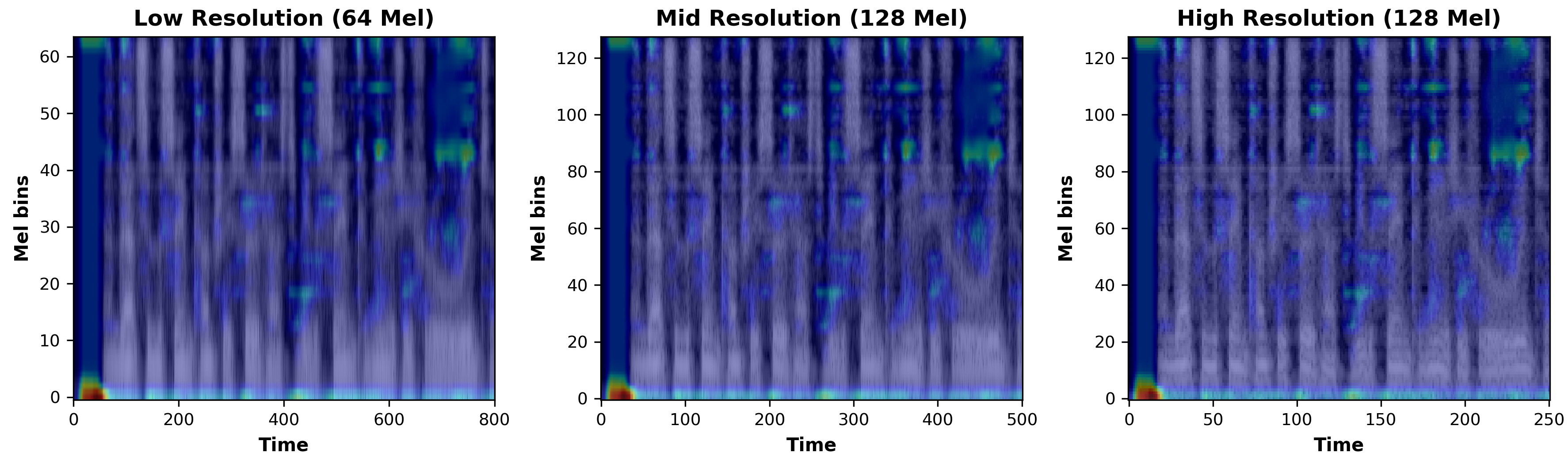}
    \caption{Grad-CAM–based interpretability analysis on the In-the-Wild audio deepfake dataset, demonstrating distributed attention across time and frequency.}
    \label{fig:interp_itw}
\end{figure}

The interpretability results for the In-the-Wild audio deepfake dataset are shown in Figure~\ref{fig:interp_itw}. In contrast to curated benchmarks, real-world samples exhibit substantial variability in speakers, background noise, compression artifacts, and recording devices.

Notably, the model does not rely on isolated spectral peaks but instead aggregates evidence over extended time spans. High-resolution activations remain informative for detecting synthesis artifacts, while mid-resolution representations capture temporal dynamics that persist across diverse recording conditions. Across all three datasets, a consistent trend emerges: no single resolution dominates the decision process. Instead, the proposed framework dynamically integrates coarse temporal context and fine spectral detail, enabling robust detection across synthesis-based, replay-based, and real-world deepfake scenarios. 

\subsection{Ablation Analysis on the FoR Dataset}

To analyze the contribution of individual components of the proposed resolution-aware framework, an ablation study is conducted on the FoR dataset as an example. Experiments are performed on three FoR variants—\textit{for-2seconds}, \textit{for-norm}, and \textit{for-rerecorded}—which progressively increase in difficulty from temporally normalized signals to replayed audio with channel distortions. All ablation models follow identical training and evaluation protocols, and performance is reported using EER and ROC-AUC.

Four model variants are evaluated:
\begin{itemize}
    \item \textbf{Full Model}: The complete proposed framework with multi-resolution inputs, cross-scale attention, and cross-resolution consistency learning.
    \item \textbf{No-Consistency}: The consistency regularization between resolutions is removed.
    \item \textbf{Single-Resolution}: Only a single Mel-spectrogram resolution is used, discarding multi-scale inputs.
    \item \textbf{No-Attention}: Cross-scale attention is replaced by simple averaging across resolutions.
\end{itemize}

Table~\ref{tab:for_ablation} summarizes the ablation results across the three FoR subsets.

\begin{table}[t]
\centering
\caption{Ablation results on the FoR dataset. Lower EER and higher AUC indicate better performance.}
\label{tab:for_ablation}
\begin{tabular}{lcccc}
\hline
\textbf{Dataset} & \textbf{Model Variant} & \textbf{EER} & \textbf{AUC} \\
\hline
\multirow{4}{*}{for-2sec}
 & Full Model & 0.0057 & 0.993 \\
 & No-Consistency & 0.0086 & 0.974 \\
 & Single-Resolution & 0.0114 & 0.962 \\
 & No-Attention & 0.0427 & 0.953 \\
\hline
\multirow{4}{*}{for-norm}
 & Full Model & 0.0046 & 0.999 \\
 & No-Consistency & 0.0071 & 0.999 \\
 & Single-Resolution & 0.0094 & 0.998 \\
 & No-Attention & 0.0124 & 0.985 \\
\hline
\multirow{4}{*}{for-rerecorded}
 & Full Model & 0.0454 & 0.9900 \\
 & No-Consistency & 0.0615 & 0.9814 \\
 & Single-Resolution & 0.0846 & 0.8970 \\
 & No-Attention & 0.261 & 0.811 \\
\hline
\end{tabular}
\end{table}

Several consistent trends can be observed across all FoR subsets. First, removing the cross-scale attention mechanism leads to the most severe performance degradation, particularly in the replayed \textit{for-rerecorded} condition. This highlights the importance of adaptive resolution weighting when spoofing cues are distorted by channel effects. Second, restricting the model to a single resolution results in a noticeable increase in EER, confirming that spoofing artifacts are distributed across complementary temporal and spectral scales. The degradation is more pronounced as acoustic conditions become less controlled. Finally, removing cross-resolution consistency learning leads to a moderate but consistent performance drop across all subsets. This suggests that consistent learning improves robustness by preventing over-reliance on resolution-specific artifacts and encouraging coherent decision-making across scales.

% -------------------------------------------------

\section{Discussion}

This study investigates the role of explicit cross-resolution modeling in improving the robustness and generalization of audio deepfake detection systems. The experimental results across controlled, replay-based, and real-world datasets consistently demonstrate that modeling interactions among multiple time--frequency resolutions provides significant advantages over conventional single-resolution or implicitly fused approaches.

Several important findings emerge from the experimental evaluation. First, the proposed resolution-aware framework achieves near-ceiling performance on controlled benchmarks such as ASVspoof 2019 LA, confirming its effectiveness in detecting algorithmic artifacts introduced by modern TTS and VC systems. More importantly, the method maintains strong performance under challenging replay-based and real-world conditions, as evidenced by the results on ASVspoof- 2019 PA, FoR rerecorded data, and the In-the-Wild Audio Deepfake dataset. These scenarios are characterized by channel distortions, background noise, and unknown spoofing methods, which often cause severe degradation in existing detectors. Second, the ablation analysis on the FoR dataset reveals that no single architectural component alone accounts for the observed performance gains. Cross-scale attention plays a critical role in dynamically weighting resolution-specific features, particularly under replay-induced distortions, while consistency learning contributes to stabilizing representations across resolutions. The degradation observed when these components are removed becomes increasingly pronounced as recording conditions become less controlled, highlighting their importance for robustness rather than merely improving performance on clean data. Third, interpretability analyses provide qualitative evidence supporting the quantitative findings. Grad-CAM visualizations indicate that the proposed model does not rely on isolated spectral peaks or narrow frequency bands. Instead, it aggregates information across time and frequency at multiple resolutions, adapting its attention depending on the acoustic context. This behavior is especially evident in replayed and in-the-wild scenarios, where artifacts are spatially dispersed and resolution-dependent.

The primary novelty of this work lies in its explicit formulation of audio deepfake detection as a cross-resolution learning problem. While prior studies have explored multi-feature or multi-scale representations, they typically rely on implicit fusion strategies such as concatenation or pooling. In contrast, this work introduces a principled resolution-aware framework that (i) explicitly models interactions among multiple spectral resolutions via cross-scale attention and (ii) enforces resolution consistency through a dedicated regularization objective. This combination encourages the model to focus on resolution-invariant cues that are more robust to channel variability and replay effects. Furthermore, the proposed architecture achieves these improvements while remaining compact and computationally efficient. With fewer than $2 \times 10^5$ trainable parameters and sub-gigaflop complexity, the model demonstrates that robustness and efficiency need not be mutually exclusive in audio deepfake detection. A key strength of the proposed method is its consistent performance across a wide spectrum of spoofing conditions, ranging from clean synthesis-based attacks to unconstrained real-world deepfakes. The use of a shared encoder ensures parameter efficiency and mitigates overfitting, while cross-scale attention enables adaptive feature integration without introducing excessive complexity. Another notable strength is the interpretability of the learned representations. The alignment between quantitative performance gains and qualitative saliency patterns provides confidence that the model learns meaningful and generalizable cues rather than exploiting dataset-specific artifacts. This property is particularly important for forensic and security applications, where trustworthiness and transparency are essential.

Despite its strengths, the proposed framework has several limitations. First, the evaluation is limited to Mel-spectrogram–based representations. Although Mel features are widely adopted and perceptually motivated, other representations, such as raw waveforms \cite{lee2025dual}, phase-based features \cite{prashnani2024generalizable}, or learned filterbanks \cite{pfister2018learning}, may capture complementary information that is not fully exploited in this work. Second, the consistency learning objective is applied only to bona fide speech under the assumption that genuine speech exhibits resolution-invariant characteristics. While empirically effective, this assumption may not hold universally across all recording conditions, particularly in highly degraded environments.

Finally, although the model generalizes well across the evaluated datasets, the rapidly evolving nature of generative speech models may introduce new artifacts that were not present during training \cite{lyu2020deepfake}. Continuous adaptation and evaluation will therefore remain necessary \cite{zhang2025audio,dinh2024towards}. Several promising directions for future work emerge from this study. Extending the resolution-aware framework to incorporate additional feature modalities, such as phase information or self-supervised speech representations, could further improve robustness. Exploring adaptive or data-driven resolution selection mechanisms may also reduce computational overhead while preserving performance. Another important direction is the integration of domain adaptation or continual learning strategies to address evolving synthesis techniques and recording conditions \cite{chen2025continual,kabir2025comprehensive}. Finally, applying the proposed framework to multimodal deepfake detection, where audio is jointly analyzed with visual or textual cues, represents a natural extension with significant practical relevance.

% -------------------------------------------------
\section{Conclusion}

This paper presented a resolution-aware audio deepfake detection framework that explicitly exploits complementary time--frequency representations through cross-scale attention and consistency learning. By jointly modeling multiple spectral resolutions, the proposed approach addresses a key limitation of existing detection systems that rely on single-resolution features or implicit fusion strategies, which often struggle to generalize under channel distortions and real-world recording conditions. Comprehensive evaluations on ASVspoof 2019, the Fake-or-Real (FoR) dataset, and an In-the-Wild audio deepfake benchmark demonstrate that the proposed method delivers strong and consistent performance across algorithmic, replay-based, and unconstrained spoofing scenarios. The framework achieves near-ceiling accuracy on controlled benchmarks while exhibiting improved robustness under replay and real-world conditions, all within a compact and computationally efficient model design.

The experimental and interpretability analyses further highlight the importance of cross-resolution attention and consistency regularization in capturing resolution-invariant spoofing cues. These components enable the model to dynamically integrate coarse temporal context with fine-grained spectral artifacts, leading to stable and generalizable representations. Overall, this work presents a principled and practical solution for robust audio deepfake detection, underscoring the value of explicit cross-resolution modeling for deployment in realistic and evolving acoustic environments.

\section*{Acknowledgment}
The author would like to thank the open research community for making the ASVspoof 2019, Fake-or-Real (FoR), and In-the-Wild Audio Deepfake datasets publicly available, which made this study possible. The author also acknowledges Grammarly and OpenAI's ChatGPT for their language refinement purposes.

\section*{Declaration of Competing Interest}
The author declares that there are no known competing financial interests or personal relationships that could have appeared to influence the work reported in this paper.

\section*{Data and Code Availability}

The datasets used in this study are publicly available from their respective sources: the ASVspoof 2019 dataset, the Fake-or-Real (FoR) dataset, and the In-the-Wild Audio Deepfake dataset. Access to these datasets is subject to the terms and conditions specified by their original providers. The code used to implement the proposed resolution-aware framework, including preprocessing, training, and evaluation scripts, will be made publicly available upon acceptance of the paper to support reproducibility and future research.

% =========================
% References
% =========================
\bibliographystyle{elsarticle-num}
\bibliography{references}

\end{document}